\documentstyle[12pt,twoside,fleqn,espcrc1,epsf,epsfig,amssymb]{article}
\newcommand{\beq}{\begin{equation}}
\newcommand{\eeq}{\end{equation}}
\newcommand{\beqa}{\begin{eqnarray}}
\newcommand{\eeqa}{\end{eqnarray}}

\newcommand{\AmS}{{\protect\the\textfont2
  A\kern-.1667em\lower.5ex\hbox{M}\kern-.125emS}}

\title{Unification of the physics of nucleons and nuclei}
\author{Ulf-G. Mei{\ss}ner\address{Universit\"at Bonn, 
HISKP (Th), Nu{\ss}allee 12-14, 
D-53115 Bonn, Germany}
}
\begin{document}

\maketitle

\begin{abstract}
\noindent I outline an ambitious program which aims to achieve
a unified description of nucleon and nuclear properties based
on {\em one} chiral effective field theory.
\end{abstract}

\section{WHAT AND WHY~?}

\noindent Tremendous progress has been made in calculating the properties of 
few--nucleon systems to a very high level of accuracy, as witnessed by many talks 
during this conference. In particular, precise calculations of few--body systems are
needed since light nuclei are  (almost) the only laboratory for extracting the elusive
properties of the neutron. However, the highly successful 
so--called standard approach has reached
certain limitations, for example two-- and three--nucleon forces are not calculated
consistently with each other and the physics underlying the nucleon structure
is treated very differently from the one leading to the nuclear bound and scattering
states. This induces some theoretical uncertainty, which is difficult to quantify.
For these and other reasons, we really would like to have a description of nucleons
{\em and} nuclei from {\em one} field theory. Such a scheme must be rooted in the
symmetries of QCD and, being based on field theory, should allow for a straightforward
implementation of gauge and other symmetries. Furthermore, we wish to have at 
our disposal some small parameter(s), leading to a systematic and controlled
expansion. Consequently, theoretical error bars take over the often (mis)used 
$\chi^2/{\rm dof} = 1$ criterion. Furthermore, such a scheme should be 
extendable to cope with inelastic 
processes and the onset of relativity. This can be formulated as a {\em new paradigm}:
To calculate nucleon as well as nuclear properties and dynamics, use quantum field
theory and organize the theory in a systematic expansion based on (a) small
parameter(s). Needless to say that the working tool will be an appropriately tailored
effective field theory (EFT). Now one might ask the question whether such a program
is not too ambitious from the beginning? 
I do not believe so, chiefly because QCD itself
offers a rich and intricate pattern of symmetries and their violations to achieve
this goal. More precisely, the spontaneous, explicit and anomalous (chiral) symmetry
breaking that QCD is supposed to undergo
can be analyzed in terms of EFT. Furthermore, the recent developments in formulating
a nuclear EFT allow for the required link between nucleon and nuclear physics, as I will show
in the following sections. In this short write-up, I can only give a flavor of
these developments but I hope that the reader will become sufficiently curious.
For more introductory material and reviews on these various intertwined topics I refer to
the ``handbook of QCD'' \cite{QCD}.
\vfill

\section{HOW~?}

\noindent 
For the light quark flavors, the QCD Hamiltonian possesses an approximate chiral symmetry.
In the chiral limit of vanishing quark masses, this symmetry is exact. It is, however, not
realized in the ground state or the particle spectrum. Rather, the chiral symmetry is
broken down to its vectorial subgroup with the appearance of three pseudoscalar Goldstone
bosons, the pions (I restrict myself here to the two flavor case). If one considers reactions
with at most one nucleon in the asymptotic in-- and the out--states, 
a perturbative expansion in terms
of small momenta and quark masses is possible. Denoting by $Q$ a generic small parameter,
any S--matrix element or transition current can be written as
\beq
{\mathcal M} = \sum_\nu {\mathcal M}_\nu~, \quad {\rm with} \quad
{\mathcal M}_\nu = Q^\nu \, f\left( Q/\lambda , g_i \right)~,
\eeq
where $\lambda$ is a renormalization scale, $g_i$ denotes coupling ({\em low--energy})
constants of the contact interactions appearing at each order,  and the
function $f$ is of order one. The important observation is that the index $\nu$ 
is bounded from below because of chiral symmetry. Thus, provided $Q$ is parametrically
small, this defines a  perturbative expansion in terms of tree and (pion) loop
graphs (see also the discussion on the structure of the nucleon below). For systems
with baryon number larger than one, an additional non--perturbative resummation is
necessary to generate large scattering lengths and shallow nuclear bound states,
as described in the talks by Beane \cite{Silas} and Epelbaum \cite{Evgeny}
at this conference.
Next, I briefly discuss the structure of the effective chiral
Lagrangian, which decomposes into the purely mesonic, the meson--nucleon and 
the nucleonic sectors. It has the generic form
\beq
{\mathcal L}_{\rm eff} = {\mathcal L}_{\pi\pi} +  {\mathcal L}_{\pi N}
 +  {\mathcal L}_{N N} + \ldots ~,
\eeq
where the ellipsis stands for terms with six or more nucleons fields. Each
of these terms has a chiral expansion, for example
${\mathcal L}_{\pi N} = {\mathcal L}_{\pi N}^{(1)} +  {\mathcal L}_{\pi N}^{(2)}
+ {\mathcal L}_{\pi N}^{(3)}  + \ldots ~,$ 
where the superscript denotes the chiral dimension. At each order, new and more
contact interactions accompanied by so--called low--energy constants (LECs) appear.
While the number of such terms quickly rises as one goes to higher orders, the
number of operators contributing to a certain process stays manageable. Consider
e.g. pion--nucleon scattering, analyzed in terms of the chiral effective pion--nucleon
Lagrangian. We have to determine 4, 5, and 4 LECs at dimension
two, three and four, respectively, from the total of 7, 23 and 118 terms at these
orders \cite{FMMS}. Furthermore, the numerical values for these LECs are understood in many
cases, rooted in  the so-called resonance saturation hypothesis (that is integrating out
heavier degrees of freedom and estimating LECs from the couplings and masses of these
heavy states),
first investigated in the meson sector \cite{EGPdR,DRV}. As an example,
take the finite dimension two coefficient $c_4$ from ${\mathcal L}_{\pi N}^{(2)}$.
From the analysis of various $\pi N$ scattering PWAs, one finds $c_4 = 3.36 \ldots
3.73\,$GeV$^{-1}$. 
In terms of resonance saturation \cite{BKMreso} (see Fig.~1), 
we get $c_4 = c_4^\Delta + 
c_4^R + c_4^\rho = (1.9 + 0.1 + 1.6)\,$GeV$^{-1} = 3.6\,$GeV$^{-1}$,
where $\Delta$, $R$ and $\rho$ denote the Delta, the
Roper resonance and the $\rho$ meson, in order. 
We thus see that not only the numerical value of this LEC can
be understood but also it comes out somewhat larger than the value based
on naive dimensional analysis (NDA), $|c_4^{\rm NDA}|\simeq 1/m\simeq 1\,$GeV$^{-1}$
(with $m$ the nucleon mass). 
For a similar investigation of the LECs in 
${\mathcal L}_{N N}$ (four--nucleon operators) see \cite{EGME}.

\parbox{8cm}{
\begin{center}
\psfig{figure=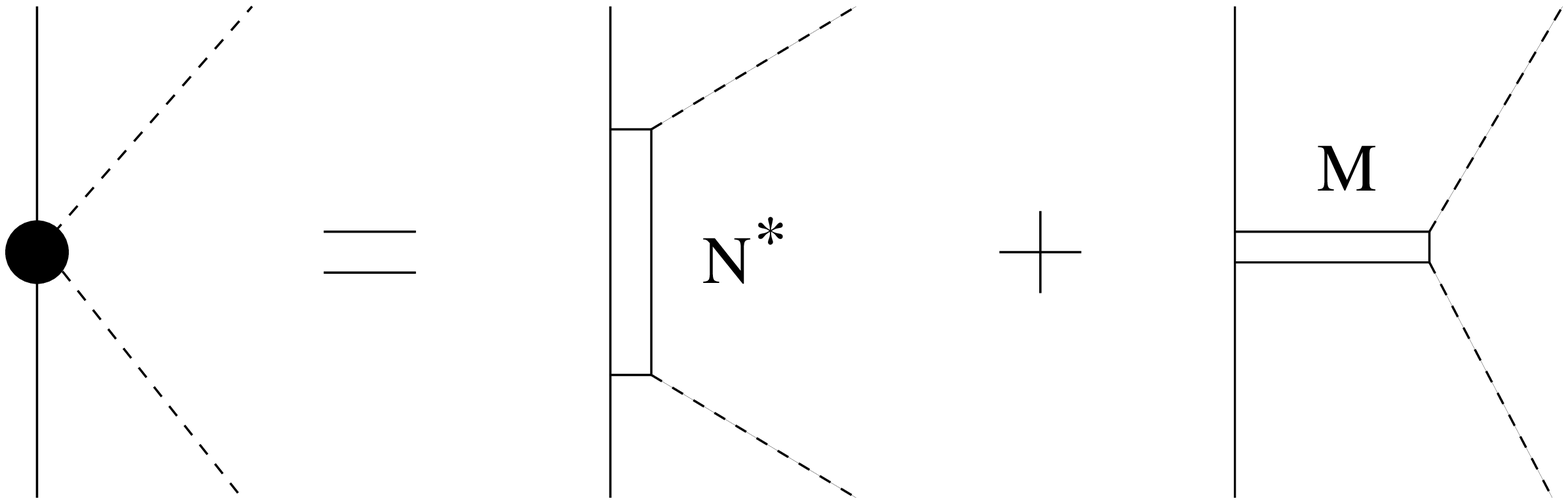,width=6cm}
\end{center}}
\hfill
\parbox{7.2cm}{\vspace{1.cm}
{\small \setlength{\baselineskip}{2.6ex} Figure~1. Resonance saturation
for a dimension two operator (filled circle). Solid and dashed lines 
stand for nucleons and pions, $N^* \, (M)$ denotes a baryon (meson)
resonance.
}}

\vspace{0.2cm}
\noindent


\section{FIRST EXAMPLE: NUCLEON AND DEUTERON ELECTROMAGNETIC FORM FACTORS}

\noindent As a first  example how such a unified approach works, we
consider elastic electron scattering off the nucleon and off deuterium.
First, we discuss the nucleon case. The nucleon matrix element of the conserved vector
current is parameterized in terms of two form factors $F_{1,2}$ (or, equivalently, the
Sachs form factors, $G_{E,M}$). 
These form factors are also a good example of how the complicated
structure of the nucleon, which is certainly an extended object, arises from the
structureless spin-1/2 fields in the chiral Lagrangian. 
There are two distinct contributions.
The first finite size effect is related to the already mentioned fact that 
the effective Lagrangian contains all terms allowed by  
PCT transformations and chiral symmetry. 
This leads to a string of terms of higher order couplings in the various sectors.
These couplings  parameterize all short distance 
physics contributing to the internal structure of the baryons. 
\setcounter{figure}{1}
\begin{figure}[h]
\parbox{.99\textwidth}{\epsfig{file= 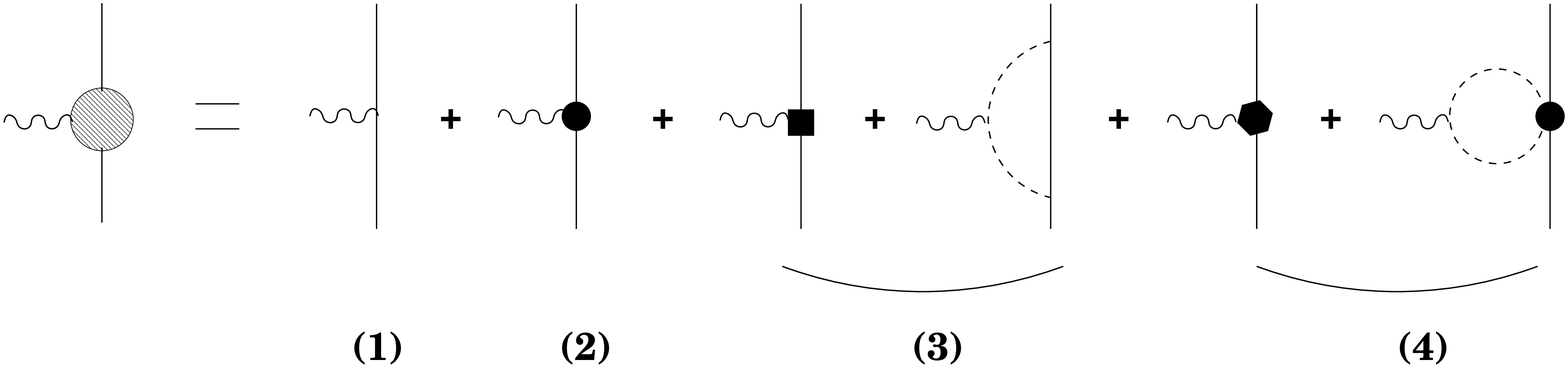,width=.90\textwidth,silent=,clip=}}
\caption{Chiral expansion of a nucleon electromagnetic form factor. The
      lowest order graph (1) with dimension one insertions only gives the charge
      of the baryon. Tree graphs with insertions from the dimension two (2), three (3)
      and four (4) effective Lagrangian are depicted by the solid circle, square
      and sextangle, in order. Pion loop graphs start at third order (3), and
      fourth order loop graphs (4) have exactly one dimension two insertion. Only
      one typical loop graph at the orders considered is shown. Solid,
      dashed and wiggly lines denote nucleons, pions and photons,
      respectively.\label{fig:ff}
}
\end{figure}

\vspace{-0.2cm}\noindent
The other finite size effect of the nucleon only starts to get generated at the 
one-loop level, as  in every quantum field theory short--lived fluctuations can occur. 
In baryon CHPT a nucleon typically 
emits a pion, this energetically forbidden $\pi$N intermediate state lives for a 
short while and then 
the pion is reabsorbed by the nucleon, in accordance with the uncertainty principle. 
This mechanism is 
responsible for the venerable old idea of the ``pion cloud'' 
of the nucleon, which in CHPT can be put on 
the firm ground of field theoretical principles. 
This is graphically depicted and
further explained in Fig.~\ref{fig:ff}. It is, however, important to stress that
the relative strength of these contributions in general depends on the choice of
renormalization scale, for a detailed discussion I refer to \cite{BHMlat}. In the
left panel of Fig.~3 I show the result for the neutron electric form factor based on 
a fourth order calculation  employing infrared regularization \cite{KM}
in comparison to modern data extracted from polarization experiments at NIKHEF,  
MAMI and MIT-Bates. There is one LEC, which can be fixed from the neutron charge
radius measured in thermal neutron--atom scattering as indicated by the dashed
line in the figure. As discussed in detail in \cite{KM}, to achieve an
equally accurate description of the other three form factors for 
$Q^2 \lesssim 0.4\,$GeV$^2$, vector mesons must be included. For these 3-point
functions, that can be done systematically and with no new tuneable
parameters. Space forbids to show the resulting form factors, the interested
reader should consult Ref.~\cite{KM}. 
\begin{figure}[t]
\parbox{.49\textwidth}{\epsfig{file= 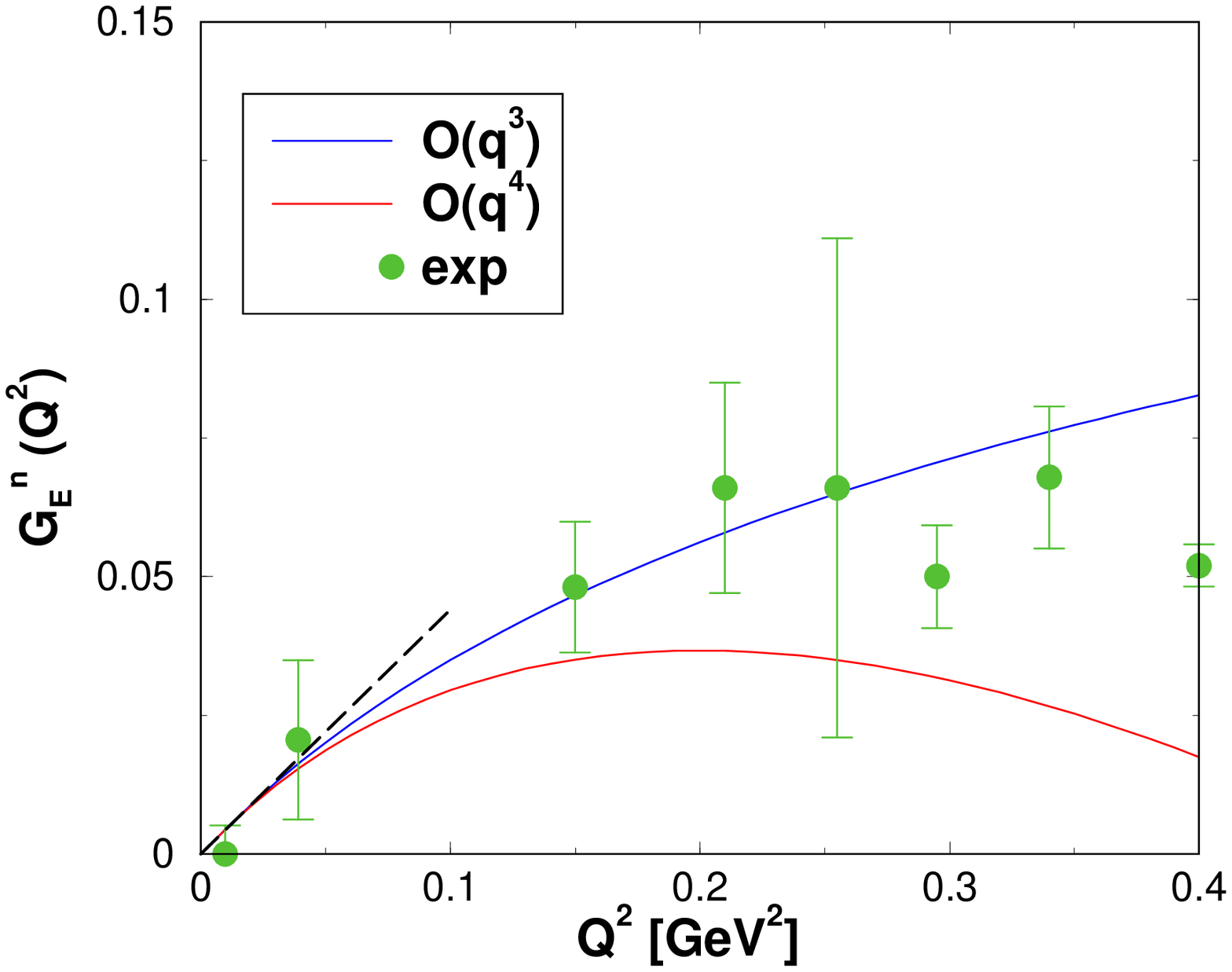,width=.47\textwidth,silent=,clip=}}\hfill
\parbox{.49\textwidth}{\epsfig{file=
    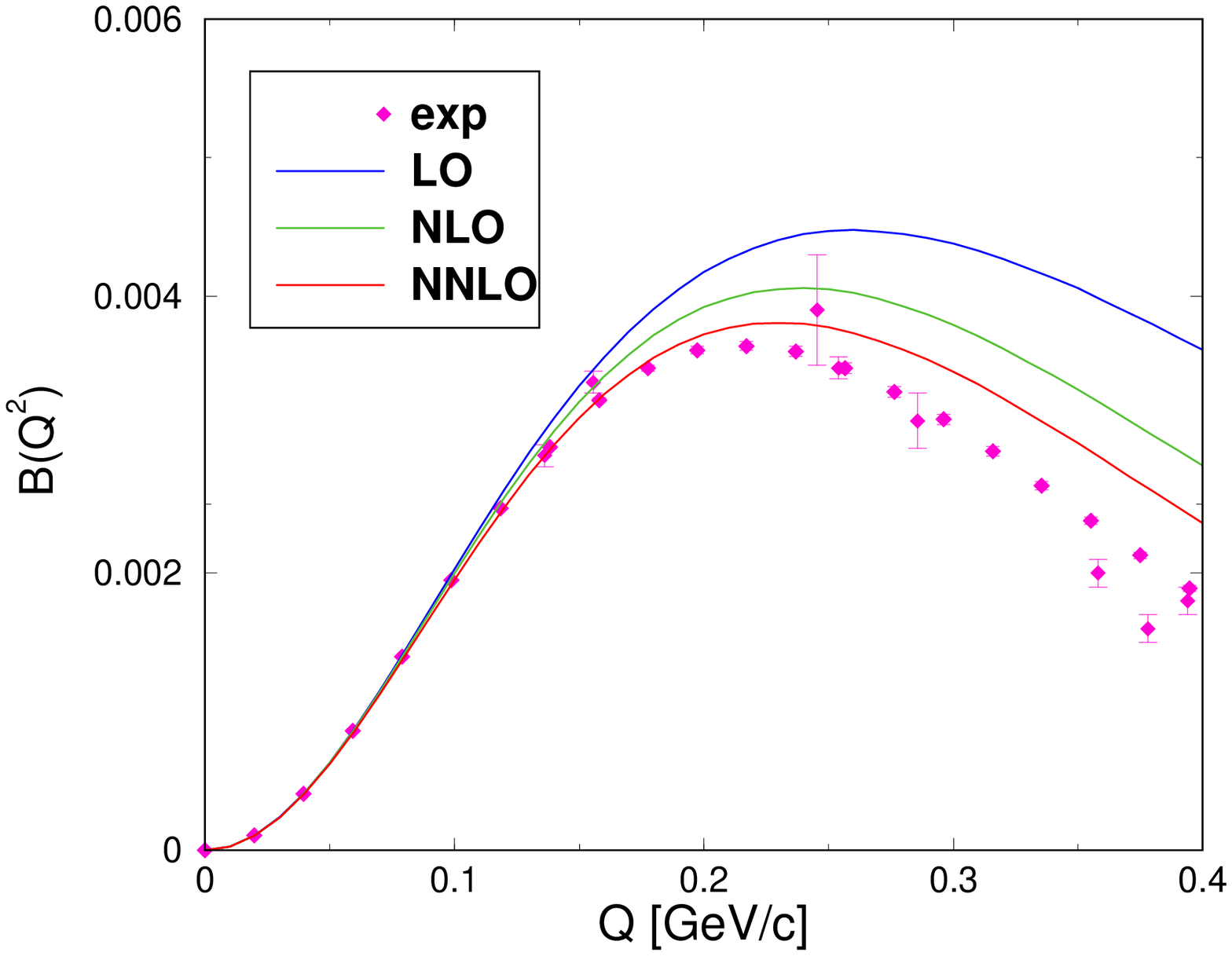,width=.47\textwidth,silent=,clip=}}
    \caption{\label{fig:sff}
           Left panel: Neutron electric form factor $G_E^n (Q^2)$ at third 
           and fourth order
           in comparison to the data, see \protect \cite{KM}. Right panel: 
           Deuteron structure function $B$ versus the photon virtuality $Q$ at 
           LO, NLO and NNLO in
           comparison to the data, see \protect \cite{WM}. 
}
\end{figure}
Let us now turn to electron scattering
off the deuteron.  The central object is  the (unpolarized) scattering cross section,
which is given in terms of two structure functions,
\beq\label{XS}
\frac{d\sigma}{d\Omega} = \left(\frac{d\sigma}{d\Omega}\right)_{\rm
  Mott} \, \left[ A(q^2) + B(q^2) \tan \frac{\theta}{2} \right]~,
\eeq
with $q^2 = -Q^2 <0$ the invariant momentum transfer squared and $\theta$ is the
scattering angle in the centre-of-mass frame. Furthermore, we have separated
the QED (Mott) cross section. These structure functions are subject to a
chiral expansion, and consist of essentially two types of contributions. The
first type of terms comprise the so-called impulse approximation contribution,  
which is nothing but the embedding
of the single nucleon form factors within the deuteron, as indicated by the
dashed boxes in the upper row of Fig.~4. In addition, there are the so--called
three--body corrections (meson exchange currents), which are parameter-free
up to NLO and have one four--nucleon--photon operator at NNLO \cite{KSWed,WM}.
The corresponding LEC can be fixed at zero momentum transfer from the deuteron magnetic
moment. In the right panel of Fig.~3 the structure function $B(Q)$ is shown at
LO, NLO and NNLO. Visibly, the description of the data improves at higher
orders and, furthermore, the corrections become smaller when going from LO to NLO to NNLO.
As stressed in \cite{WM}, the limiting factor in the accuracy
to be achieved for the deuteron structure functions stems from the description
of the single nucleon form factors. For a somewhat different approach, see 
\cite{Dan}.

\parbox{9cm}{
\begin{center}
\psfig{figure=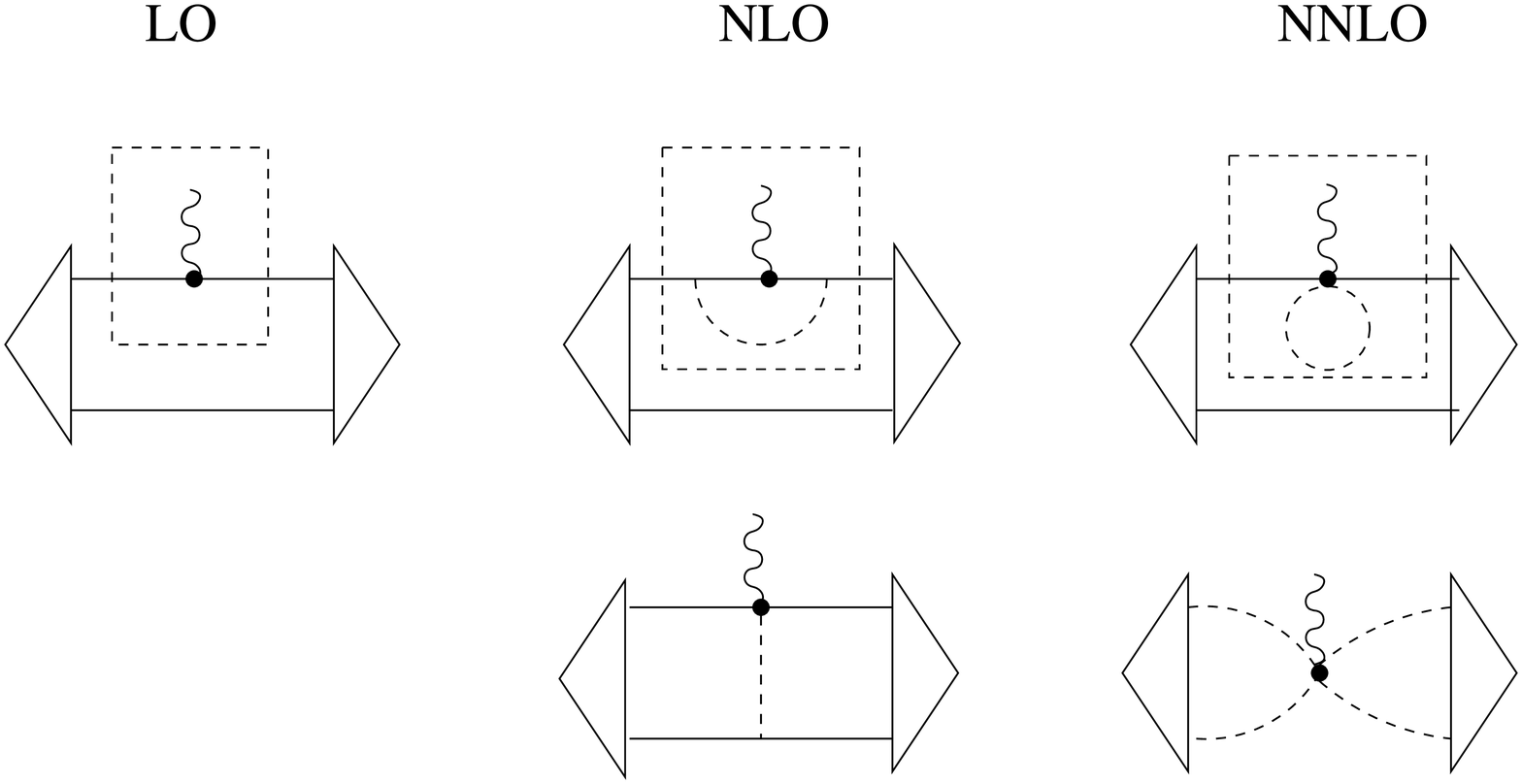,width=8.9cm}
\end{center}}
\hfill
\parbox{6.2cm}{\vspace*{.4cm}
{\small \setlength{\baselineskip}{2.6ex} Figure~4. Chiral expansion of
$ed$ scattering. The upper row refers to the single scattering contributions
(impulse approximation), the lower row to the three--body corrections (meson
exchange currents). The triangle symbolizes the deuteron wave function, for
further notation see Fig.~2. Only one representative diagram 
is shown at each order.
}}

\bigskip
\noindent While not very spectacular, I think that this example nicely shows
how indeed the nucleon and nuclear properties can arise from one effective 
field theory and it will be of utmost importance to extend this program to the
investigation of the electromagnetic structure of light nuclei.

\section{SECOND EXAMPLE: PION--NUCLEON AND PION--DEUTERON\\ SCATTERING}

\noindent
Another interesting example concerns pion--nucleon scattering at zero
energy, i.e. the S-wave scattering lengths $a^+$ (isoscalar) and
$a^-$ (isovector). Already Weinberg pointed out many decades ago that these
observables allow for a crucial test of chiral symmetry breaking. In the
mean time, the chiral expansion of these quantities has been worked out
to fourth order \cite{BKMpin,FM4} (for brevity, we do not display the
fourth order contribution to $a^+$ but refer the reader to \cite{FM4})
\beqa
4\pi (1+\mu) \, a^+ &=& \frac{M_\pi^2}{F_\pi^2}\left( \Delta - 
\frac{g_A^2}{4m} \right) + \frac{3 g_A^2 M_\pi^3}{64 \pi F_\pi^4}
+ {\mathcal O}\left( M_\pi^4 \right)~, \\
4\pi (1+\mu) \, a^- &=&  \frac{M_\pi}{2 F_\pi^2} +  
\frac{4M_\pi^3}{F_\pi^2} \left( \bar D + \frac{g_A^2}{32m^2} \right)
 + \frac{M_\pi^3}{16 \pi^2 F_\pi^4}
+ {\mathcal O}\left( M_\pi^5 \right)~, 
\eeqa
where $\mu = M_\pi/m \simeq 1/7$ is the small threshold parameter,
 $ \Delta = -4c_1 + 2c_2 + 2c_3$ and $ \bar D = \bar d_1 + 
 \bar d_2 +  \bar d_3 + 2\bar d_5$ are combinations of dimension two and three
LECs, respectively. These can be determined by a fit to the existing PWAs
at finite energy, so that one can predict the scattering lengths and the
phases at higher energies. This leads to the results collected in 
Table~\ref{tab:as} \cite{FM4}. We see that for $a^-$ the convergence is
good and the value has a small theoretical uncertainty (the low value from Fit~2 
should be excluded as discussed in detail in \cite{FM3}). On the other hand, the
convergence for $a^+$ is moderate and the theoretical error quite large.  
\begin{table}[t]
\begin{center}
  \caption{Convergence of the S--wave scattering lengths for CHPT
   fits to the Karlsruhe (Fit 1), Z\"urich (Fit 2), and VPI (Fit 3)
    phase-shift analyses of $\pi N$ data. 
   $O(p^n)$ means that all terms up-to-and-including
   order $n$ were included. 
   Units are $M_\pi^{-1}$. 
    \label{tab:as}}
\vspace{0.2cm}
\begin{tabular}{|r|r|r|r|r|}
    \hline
          & $O(p)$  &   $O(p^2)$  &  $O(p^3)$ 
          & $O(p^4)$  \\
    \hline\hline
           Fit 1      & 0.0  & 0.0046  & $-$0.0100    & $-$0.0096 \\
$a^+ \quad$ Fit 2    & 0.0  & 0.0024  &    0.0049  &    0.0045 \\
           Fit 3      & 0.0  & 0.0101  &    0.0014  &    0.0027 \\ 
\hline
           Fit 1      & 0.0790 & 0.0790  &  0.0905  &    0.0903 \\
$a^- \quad$ Fit 2     & 0.0790 & 0.0790  &  0.0772  &    0.0771 \\
           Fit 3      & 0.0790 & 0.0790  &  0.0870  &    0.0867 \\
\hline
  \end{tabular}
\vspace{-0.5cm}
\end{center}\end{table}
These results are consistent with the precise pionic hydrogen data
obtained at PSI, which again do not allow to pin down $a^+$ very precisely
(see the discussion in \cite{BBEMP}). It is also important to stress that
only with the recent work of the Bern group \cite{Bern} we really have
a clear-cut analysis of the isospin breaking effects that are so important
to properly deduce the strong contribution to the energy shift of the
$\pi^- p$ atom (note also that the conventional methods of subtracting the
electromagnetic effects from low energy $\pi N$ scattering data has been put in doubt
recently \cite{FMiso}).  
\setcounter{figure}{4}
\begin{figure}[htb]
\parbox{.99\textwidth}{\begin{center}
\epsfig{file= 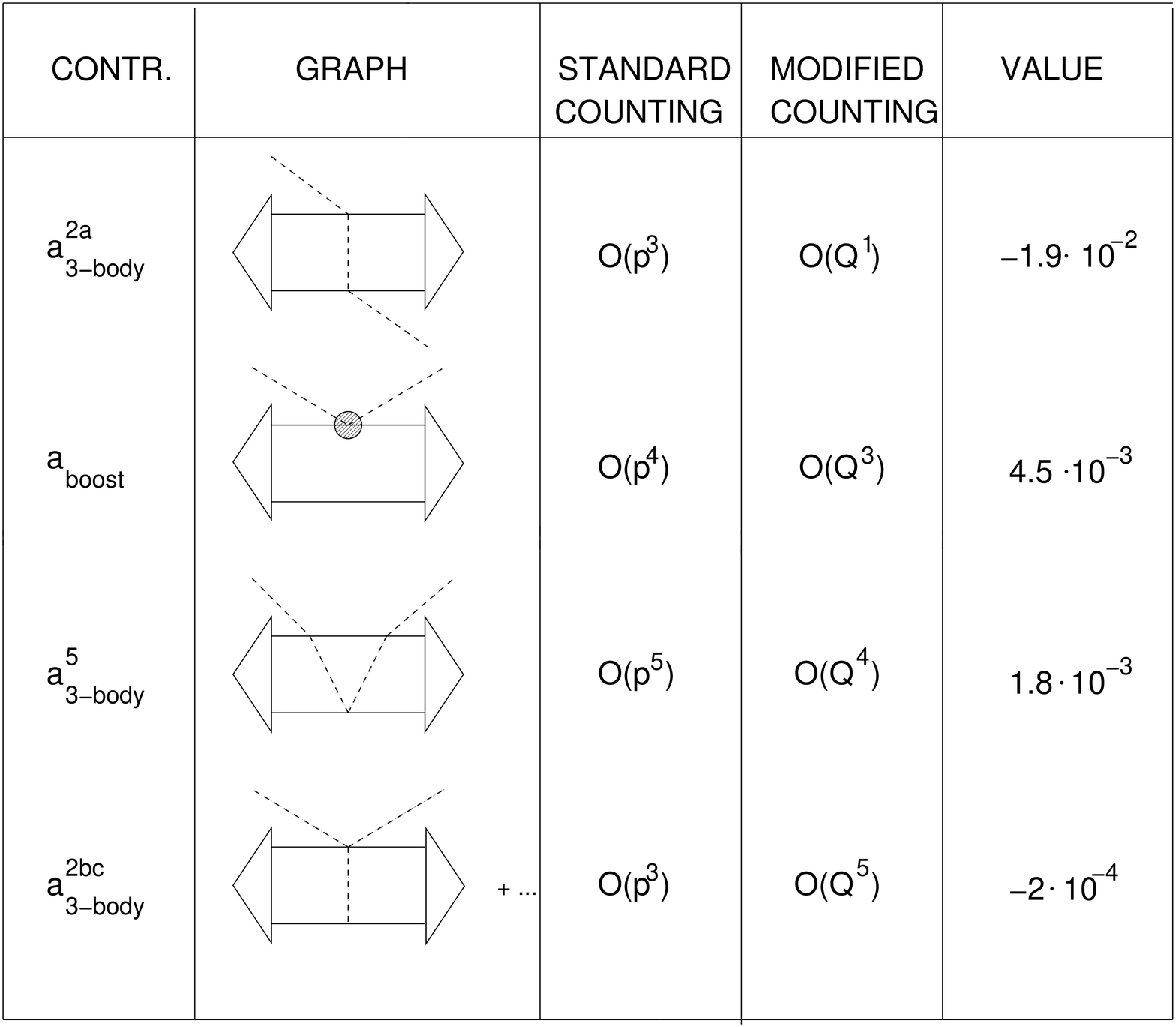,width=.72\textwidth,silent=,clip=}\end{center}}
\vspace{-0.5cm}
\caption{Various contributions to the real part of the $\pi d$ scattering length 
         as explained in the text. The filled circle denotes an insertion from 
         ${\mathcal L}_{\pi N}^{(2)}$.
     \label{fig:pidscheme}
}
\end{figure}
On the other hand, the deuteron is an isoscalar target
and thus to leading order entirely sensitive to $a^+$. Beyond leading
order, rescattering and other type of three--body corrections play a role,
and these are also sensitive to the isovector scattering length. This
problem has already been addressed to third order in Refs.~\cite{Weinpid,BBLMvK}
and to fourth order in \cite{BBEMP}. Surprisingly, at fourth order there
is only one novel type of correction 
related to boosting the single scattering amplitudes (cf. the contribution
$a_{\rm boost}$ in Fig.~5) ,
all three--body corrections at this order vanish (for details see \cite{BBEMP}).
However, as it is known since long, the rescattering contribution (the graph
$a^{2a}_{\rm 3-body}$ in Fig.~5) is much larger than the other terms at this 
order (the graphs $a^{2bc}_{\rm 3-body}$ in Fig.~5), see e.g. \cite{RW}. 
On the other hand, in the standard CHPT counting, numerically large contributions
like e.g. multiple rescattering (the  graph $a^5_{\rm 3-body}$ in Fig.~5) are
suppressed. Consequently, one has to deal in a systematic fashion with this
issue. This was done in \cite{BS}, were it was shown that the rescattering type
graphs are IR enhanced due to the small binding momentum in the deuteron,
\beq
\gamma = \sqrt{m_D \, E_D / 2} = 45~{\rm MeV} \ll M_\pi \simeq 140~{\rm MeV}~,
\eeq
with $m_D \,(E_D)$ the deuteron mass (binding energy). More precisely, whenever
the exchanged pions can go on (close to) their mass shell, as it is the
case for zero energy scattering, one has an additional scale to cope with.
This can be accounted for by a modified power counting, where one counts typical
momenta in the deuteron as order $p^2$, with $p$ the generic small parameter 
of CHPT. Amplitudes with powers of $M_\pi$ are then suppressed and, denoting
by $Q$ the small parameter in the modified counting, one obtains an ordering
of the various contributions that is consistent with the numerical
evaluation of the different mechanisms, as depicted in Fig.~5.
Within this new scheme, there are only a few more diagrams to calculate. 
At ${\mathcal O}(Q^3)$, one has graphs like $2a$ with either two dimension
two or one dimension three (contact term and loop) insertion. At  
${\mathcal O}(Q^4)$, the only new contribution stems from $a^5_{\rm 3-body}$, cf.
Fig.~5. With that, the final formula for the real part of 
$\pi$d scattering length, valid to $O(Q^4)$ in the
modified power counting, is
\begin{eqnarray}
{\rm Re}\,a_{\pi d}&=&2\frac{(1+\mu)}{(1+\mu /2)}\,\left( a^+ \, +\, 
(1+\mu )\Big\lbrack (a^+)^2-2(a^-)^2 \Big\rbrack 
\frac{1}{2{\pi^2}}\Bigg\langle 
\frac{1}{{\vec q\,}^{\, 2}}\Bigg\rangle_{\sl wf}
\qquad\qquad\qquad\qquad \right. \nonumber\\
&& \left.  
+(1+\mu )^2\Big\lbrack (a^+)^3-2(a^-)^2(a^+-a^-) \Big\rbrack 
\frac{1}{4{\pi}}\Bigg\langle 
\frac{1}{|{\vec q}\, |}\Bigg\rangle_{\sl wf}\, \right) 
\,+\, {a_{\rm boost}}\, +\, O(Q^5)\ .
\label{thefullthing}
\end{eqnarray}
where we have  subsumed higher order effects into the 
$\pi N$ scattering lengths and $\langle \dots\rangle_{\sl wf}$ 
denotes a momentum space deuteron matrix element. 
Notice that since local two--pion--four--nucleon
operators are not enhanced in the modified power counting they appear
at the same order as in baryon CHPT, namely fifth order, and do
not affect the fourth-order result given in Eq.~(\ref{thefullthing}).
The $\pi^- d$ atomic level shift has been measured to high precision 
at PSI \cite{Hauser}, yielding the complex--valued scattering length 
$a_{\pi d}=[-0.0261 (\pm 0.0005)\;+\; i\; 0.0063 (\pm 0.0007)\;] {M_\pi^{-1}}$.
One thus can analyze Eq.~(\ref{thefullthing}) using 
the real part of this number in the
$(a^+ - a^-)$--plane, as shown by the red bands in Fig.~6. This theoretical
uncertainty is due to the NLO EFT wave functions employed. If one uses the
most recent and systematic determination of the pionic hydrogen level shift
\cite{Bern} (the green band in Fig.~6) together with the empirical constraint 
from the hydrogen width shown by the grey band in Fig.~6
(which will be affected by further experimental \cite{DG} as well as theoretical
corrections), one can determine the S--wave $\pi N$ scattering lengths
\begin{equation}
a^{-} = [0.0936 (\pm 0.0011)\;] {M_\pi^{-1}}\ ; \quad
a^{+} = [-0.0029 (\pm 0.0009)\;] {M_\pi^{-1}}\ .
\label{avalues}
\end{equation}
If one instead were to use the blue band for the hydrogen shift (based on incomplete
potential model calculations), these numbers would read
$a^{-} =[0.0917 (\pm 0.0013)\;] {M_\pi^{-1}}\,$ and
$a^{+} = [-0.0034 (\pm 0.0008)\;] {M_\pi^{-1}}$.
The theoretical error bars from both determinations should be considered somewhat 
optimistic given the  neglect of isospin violation in \cite{BBEMP}.
For a comparison to other (more model--dependent) determinations and a discussion
of the imaginary part of the $\pi d$ scattering length the reader is referred to
 \cite{BBEMP}. For a first attempt towards a systematic calculation of
 pion-$^3$He scattering see \cite{BHHN}.

\parbox{9cm}{
\begin{center}
\psfig{figure=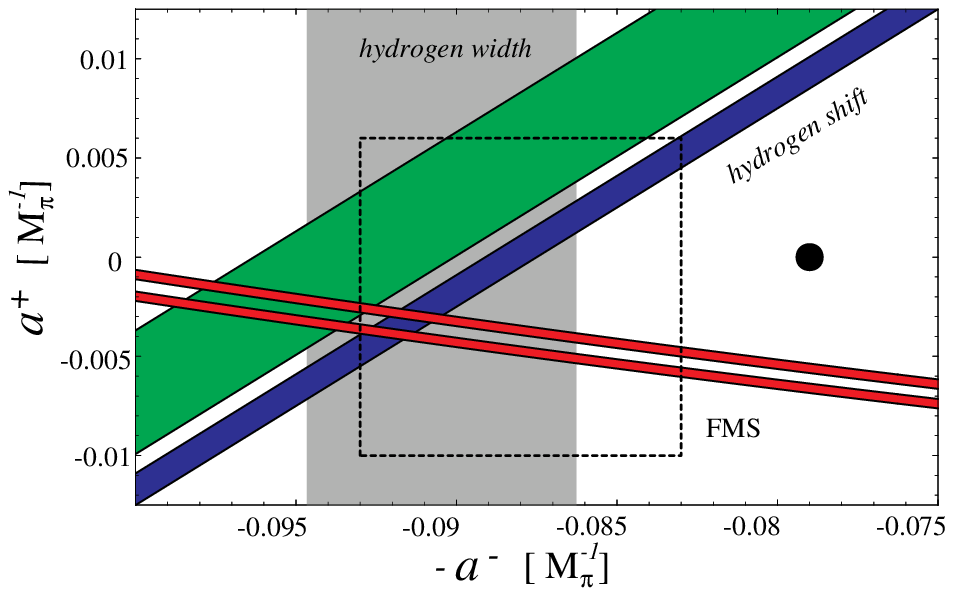,width=8.9cm}
\end{center}}
\hfill
\parbox{6.2cm}{\vspace*{1.5cm}
{\small \setlength{\baselineskip}{2.6ex} Figure~6. Determination of the
S--wave $\pi N$ scattering lengths from pionic hydrogen and deuterium
as explained in the text. The heavy dot refers to the current algebra
prediction and the black box to the CHPT determination from scattering 
data \protect \cite{FM3,FM4}.
}}

\section{SUMMARY AND OUTLOOK}

I have outlined a program that allows to calculate nucleon and nuclear
properties from one chiral effective field theory, based on a systematic
power counting, embodying the symmetries of QCD and allowing for the estimation
of theoretical uncertainties. As examples, I have 
discussed electron scattering off nucleons and deuterium and also
pion--nucleon and pion--deuteron scattering lengths, both cases displaying
an intricate interplay between chiral nucleon and chiral nuclear dynamics. Other 
examples, which have gained some popularity over the last decade, are 
pion photo/electroproduction (see e.g. the discussion in \cite{MBEG}
as well as \cite{BKMr} and \cite{BBLMK})
and Compton scattering off nucleons \cite{BKKM} and light
nuclei \cite{BMMPK}, to name just two.  
It remains to be seen how far this program can be carried out in
terms of mass number $A$ (the extension to systems with $A=3,4$ is under way), 
photon virtuality $Q^2$ or pion energy $\omega$. One also might ask at what
point other active degrees of freedom (the delta, vector mesons, $\ldots$) 
have to be taken into account explicitely? 
Many pioneering calculations have been done but much more work is required 
to really develop this scheme into a true working tool for {\em precision nuclear,
hadronic and few--body physics}.

\section*{Acknowledgements}
\noindent
I thank the organizers for the invitation and their superb work and my collaborators,
in particular Silas Beane, V\'eronique Bernard, Evgeny Epelbaum and Bastian Kubis,
for sharing their insight into the various topics discussed here.

\end{document}